\documentclass[%
longbibliography,
 reprint,
superscriptaddress,
 amsmath,amssymb,
 aps,
floatfix,
]{revtex4-1}
\usepackage[english]{babel}
\usepackage{comment}
\usepackage{natbib}
\usepackage{ulem}
\usepackage{graphicx}% Include figure files
\usepackage{dcolumn}% Align table columns on decimal point
\usepackage{bm}% bold math
\usepackage{amsmath}
\usepackage{siunitx}
\usepackage{listings}
\usepackage{mathtools}
\usepackage{float}
\usepackage{xcolor}
\usepackage{url}
\usepackage{hyperref}
\hypersetup{
    colorlinks=true,allcolors=blue,}

\begin{document}
\preprint{APS/123-QED}

\title{Positional stability of skyrmions in a racetrack memory with notched geometry}
% AUTHORS % % % % % % % % % % % %
\author{Md Golam Morshed}
\email{mm8by@virginia.edu}
	\affiliation{
		Department of Electrical and Computer Engineering, University of Virginia, Charlottesville, VA 22904, USA\looseness=-1}
\author{Hamed Vakili}
\email{hv8rf@virginia.edu}
	\affiliation{
		Department of Physics, University of Virginia, Charlottesville, VA 22904, USA\looseness=-1}
\author{Avik W. Ghosh}
	\affiliation{
			Department of Electrical and Computer Engineering, University of Virginia, Charlottesville, VA 22904, USA\looseness=-1}
	\affiliation{
	    Department of Physics, University of Virginia, Charlottesville, VA 22904, USA\looseness=-1}

\date{\today}% It is always \today, today,

\begin{abstract}
Magnetic skyrmions are chiral spin textures with attractive features, such as ultra-small size, solitonic nature, and easy mobility with small electrical currents that make them promising as information-carrying bits in low-power high-density memory, and logic applications. However, it is essential to guarantee the positional stability of skyrmions for reliable information extraction. Using micromagnetic simulations for the minimum energy path (MEP), we compute the energy barriers associated with stabilizing notches along a racetrack. We vary material parameters, specifically, the strength of the chiral Dzyaloshinskii-Moriya interactions (DMI), the notch geometry, and the thickness of the racetrack to get the optimal barrier height. We find that the reduction of skyrmion size as it squeezes past the notch gives rise to the energy barrier. We find a range of energy barriers up to $\sim 45~k_BT$ for a racetrack of $5~nm$ thickness that can provide years long positional lifetime of skyrmions for long-term memory applications while requiring a moderate amount of current ($\sim 10^{10}~A/m^2$) to move the skyrmions. Furthermore, we derive quasi-analytical equations to estimate the energy barrier. We also explore other pinning mechanisms, such as a local variation of material parameters in a region, and find that notched geometry provides the highest energy barrier. Our results open up possibilities to design practical skyrmion-based racetrack geometries for spintronics applications.
\end{abstract}
\pacs{Valid PACS appear here}
\maketitle 
\section{Introduction}
Conventional memory technology is bottlenecked by the delay in fetching instruction sets between the logic cores and the memory elements that are slower to scale. Magnetic skyrmions~\cite{skyrmion1, skyrmion2,self-focus} bear non-trivial properties, such as topological protection~\cite{topology}, ultra-small size~\cite{Caretta}, high-speed~\cite{speed}, current-induced motion including nucleation and annihilation~\cite{Sampaio,Romming}. Therefore, they have emerged as potential candidates for digital and analog information-carrying bits in low-power high-density, fast, all-electronic non-volatile memory, and logic applications~\cite{Review1,Review2,hamed1,hamed2,logic1,logic2,Review3}. Skyrmions originate and stabilize by the Dzyaloshinskii–Moriya interaction (DMI)~\cite{DMI1,DMI2} in systems lacking inversion symmetry while tuning the DMI~\cite{yang,yassine1,golam,yassine2} controls skyrmion size~\cite{size} and overall stability.
%%%%%%%%%%%%%%%%%%%%%%%%%%%%%%%%%%%%%%%%%%%%%%%%%%%%%%%%%%
%%%%%%%%%%%     Fig 1      %%%%%%
%%%%%%%%%%%%%%%%%%%%%%%%%%%%%%%%%%%%%%%%%%%%%%%%%%%%%%%%%%
\begin{figure*}[!htbp]
\includegraphics[width=.32\textwidth]{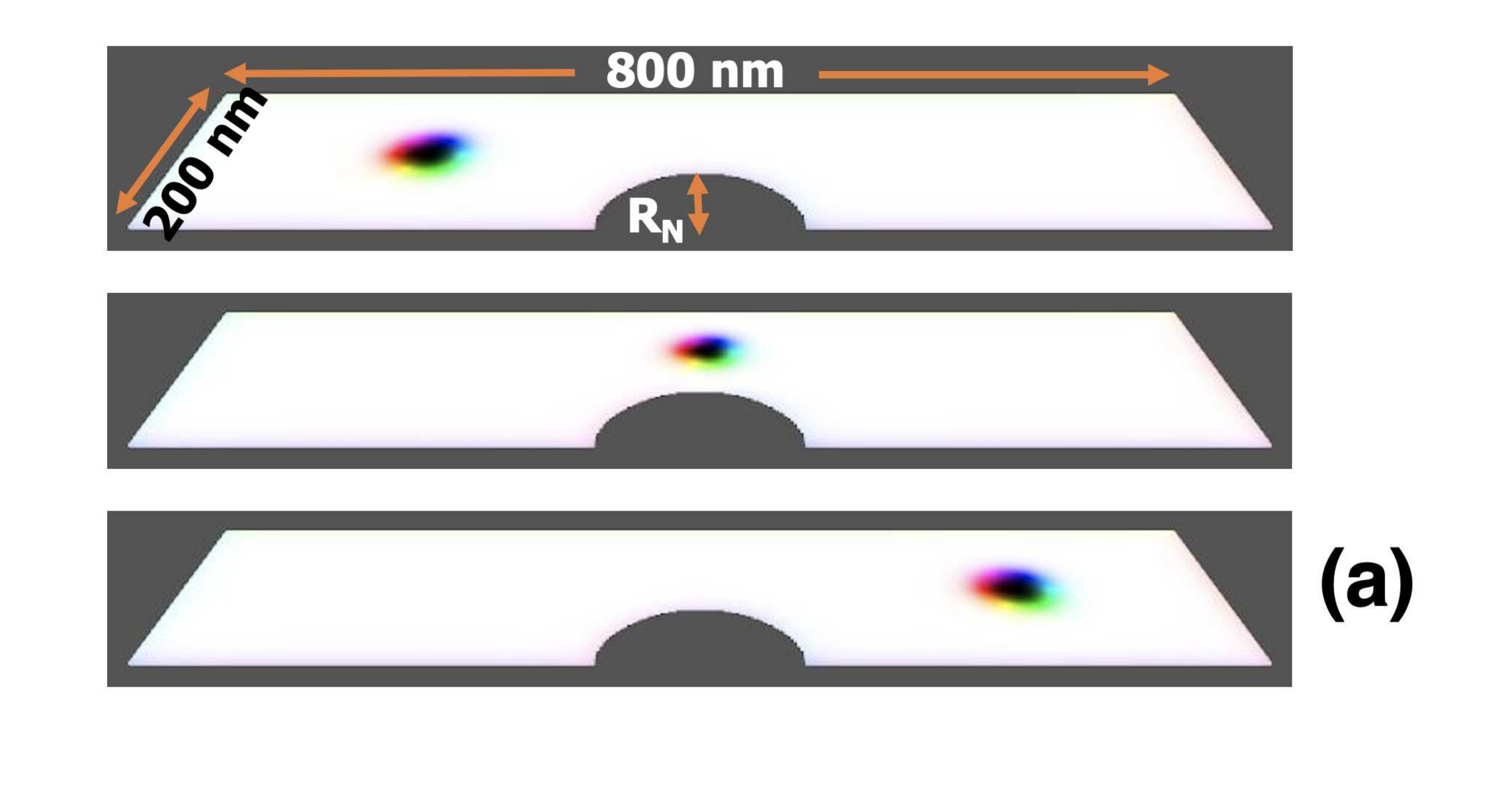}
\includegraphics[width=.32\textwidth]{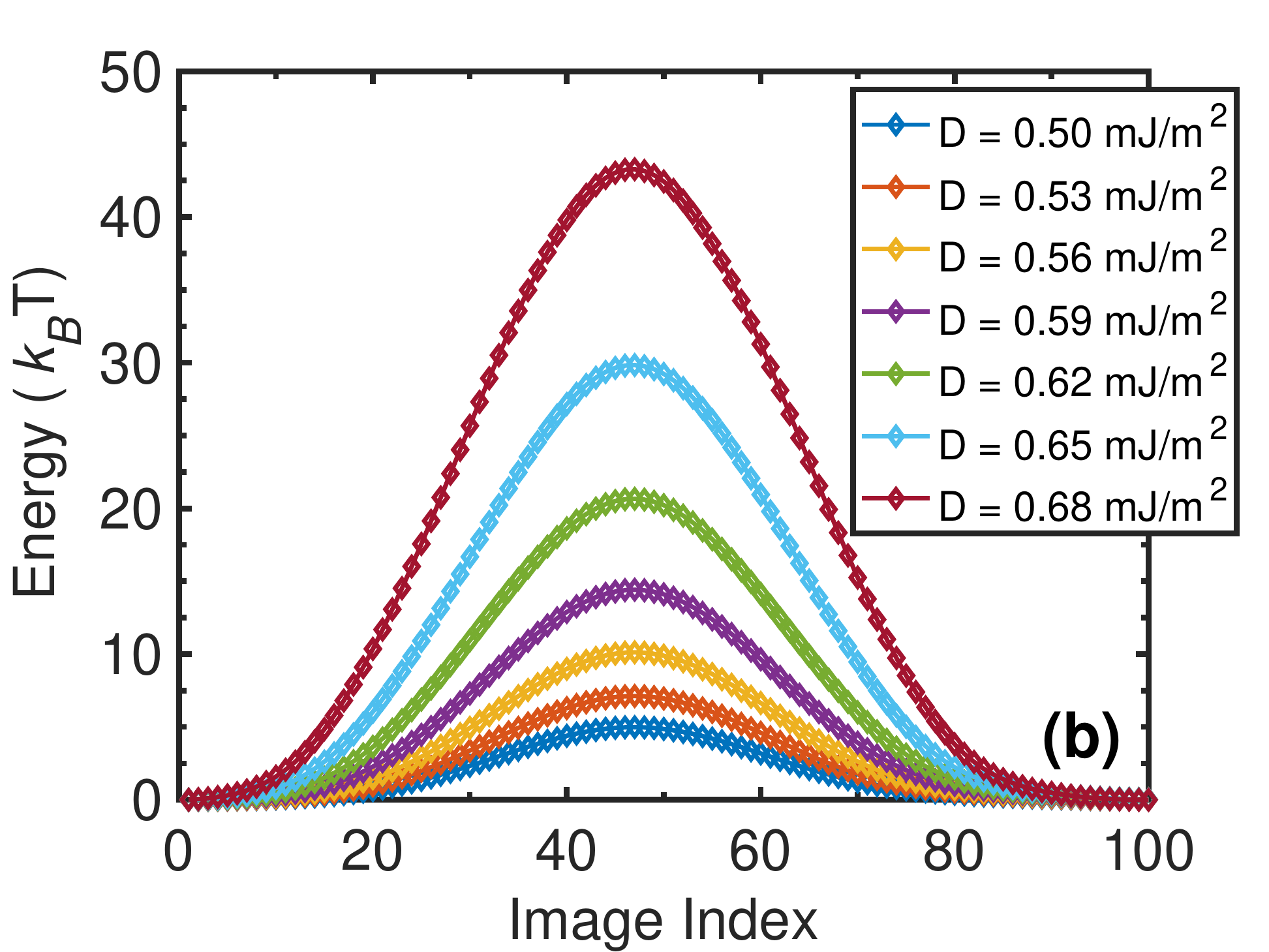}
\includegraphics[width=.32\textwidth]{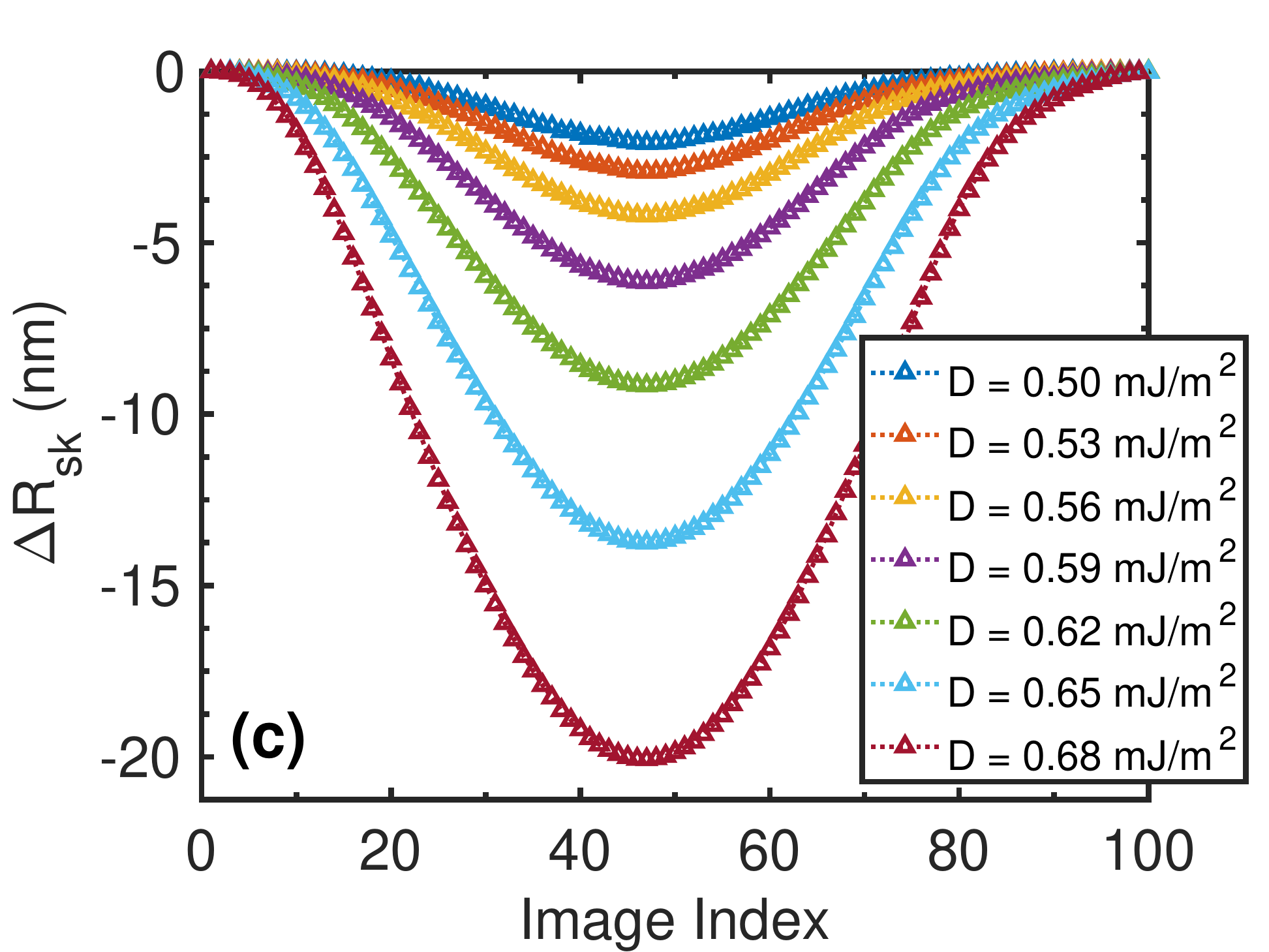}
    \caption{(a) Simulated racetrack with notched geometry. Each of the figures represents different snapshots of the skyrmion trajectory along the racetrack, referred to as `image index'. (b) The energy vs image index for the optimal trajectory of a skyrmion with varying DMIs in a $800~nm \times  200~nm$ racetrack with notch radius, $R_N = 100~nm$. The left valley, peak, and right valley correspond to the image index shown in the top, middle, and bottom panels of Fig.~\ref{fig1}(a) respectively. (c) Change in skyrmion radius, $\Delta R_{sk}$ when the skyrmion passes over the notch. Higher DMIs initiate larger initial skyrmions that undergo bigger shrinkage and larger energy costs when forced through the constriction.}
\label{fig1}
\end{figure*}\\
\indent Racetrack memory~\cite{racetrack} is one of the common platforms studied in the context of skyrmionics~\cite{skyrmion-on-the-track,Tomasello,hamed1}. In a skyrmion-based Boolean racetrack memory, information is encoded by the presence (bit “$1$”) and absence (bit “$0$”) of skyrmions at a particular position. For an analog domain utilization of a skyrmion racetrack, such as a native temporal memory for race logic \cite{hamed1}, the information is encoded directly into the spatial coordinates of the skyrmions that can be translated back into the timing information of wavefront duty cycles carrying out the race logic operations~\cite{racelogic}. The positional stability of skyrmions is a critical issue for both of these applications because a randomly displaced skyrmion can alter the bit sequence in Boolean memory applications and change the spatial coordinates hence the encoded analog timings in race logic applications. For reliability, it is essential to guarantee the positional stability of skyrmions for a certain amount of time. For instance, for long-term memory applications, it requires positional stability of years, while for cache memory, hours or even minutes would be sufficient. In an ideal racetrack, skyrmions are susceptible to thermal fluctuations and exhibit Brownian motion leading to diffusive displacement~\cite{thermal-stability,Brownian}. Moreover, skyrmions show inertia-driven drift shortly after removing a current pulse rather than stopping immediately. One way to control such undesirable motion is by engineering confinement barriers such as defects created by local variations of material parameters and notches etched into the racetrack, which ensures the pinning of skyrmions~\cite{Sampaio,pinning}.\\
\indent The interactions and dynamics of skyrmions with defects and other pinning sites have been studied over the past few years~\cite{pinning,capturing_hole,atomic_defect,pinning_review}. Few of the studies have discussed the energy barrier associated with the pinning sites peripherally~\cite{MEP1,MEP_sci}. Recently, notches have been used to achieve positional stability in domain wall-based artificial synapses~\cite{Incorvia}. Nonetheless, what is missing is a systematic analysis of a notched racetrack, the mechanics of the energy barrier, and its impact on skyrmion mobility, stability, and unintended nucleation and annihilation which sets its operating limits. The combination of required positional stability and operational current range defines an optimal `Goldilocks' regime in parameter phase space, which is the focus of this work.\\
\indent In this paper, we systematically analyze our ability to produce and tune the energy barrier in a skyrmion-based racetrack with notched geometry (Fig.~\ref{fig1}) using micromagnetic simulations. In particular, we vary the material parameters DMI (varying skyrmion sizes), the geometry of the notches (Fig.~\ref{fig2}a), and thickness of the racetrack (Fig.~\ref{fig2}c) to achieve a high tunability of the energy barrier for long-term positional stability of skyrmions. We demonstrate that the energy barrier is attributed to the constriction in the skyrmion sizes arising from the notch created in the racetrack. Additionally, we come up with an empirical equation based on our simulations (Fig.~\ref{fig3}). Furthermore, we explain and compare our simulated data with the analytical energy equations of skyrmions on an unconfined infinite plane plus a phenomenological confinement correction that show an excellent match (Fig.~\ref{fig4}a). The quantitative difference between the energy of skyrmions on an unconfined infinite plane and our simulated data is attributed to the different geometric boundary conditions (Fig.~\ref{fig4}b). We also explore other pinning sites such as local variations of material parameters to put the notched geometry into perspective with other types of defects (Fig.~\ref{fig5}). Finally, we show that the required unpinning current is small enough for skyrmion-based devices to be integrated with electrical circuits (Fig.~\ref{fig6}). Our results provide a path forward towards practical, reliable skyrmion-based racetrack memory applications.
\section{Methods}
We perform the simulations using MuMax3~\cite{mumax3}, a micromagnetics simulator that solves the Landau–Lifshitz–Gilbert (LLG) equation. The dimensions of the racetrack are length $L = 800~nm$, width $W = 200~ nm$, and thickness $t_F = 5~nm$. The simulation mesh is divided into $400 \times 100 \times 1$ grids with a cell size of $2~nm \times 2~nm \times 5~nm$ without considering periodic boundary condition. We use GdCo material parameters such as exchange stiffness $A_{ex} = 7~pJ/m$, anisotropy $K_u = 50~kJ/m^3$, saturation magnetization $M_s = 100~kA/m$ throughout all the calculations unless otherwise specified~\cite{skm_theory,Caretta,Review2}. We use varying DMIs to control the size of skyrmions because the DMI can be easily tuned by interface engineering~\cite{yassine1,golam}. We calculate the minimum energy path (MEP) using the String method~\cite{string2,String}. The basic idea of the string method is to find the transition path by evolving a curve (string) connecting two endpoints along the energy landscape and the reparametrization of the string by interpolation~\cite{string2,String}. It is an iterative method that continues until the path converges to the MEP with the desired accuracy. In our simulations, we use $100$ iterations to calculate the MEP.
\section{Results and Discussion}
%%%%%%%%%%%%%%%%%%%%%%%%%%%%%%%%%%%%%%%%%%%%%%%%%%%%%%%%%%
%%%%%%%%%%%     Fig 2       %%%%%%
%%%%%%%%%%%%%%%%%%%%%%%%%%%%%%%%%%%%%%%%%%%%%%%%%%%%%%%%%%
\begin{figure*}[!htbp]
\includegraphics[width=.32\textwidth]{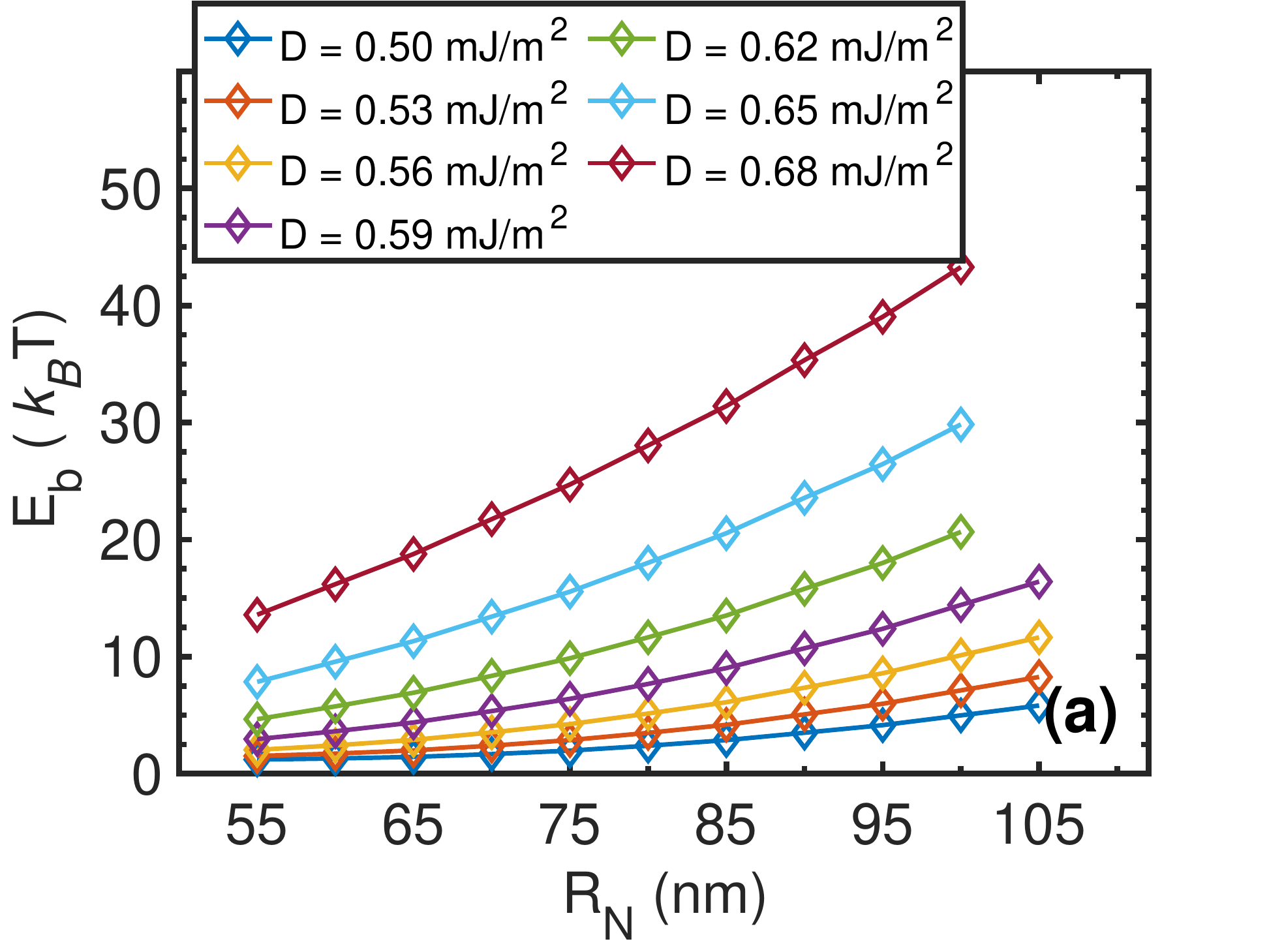}
\includegraphics[width=.32\textwidth]{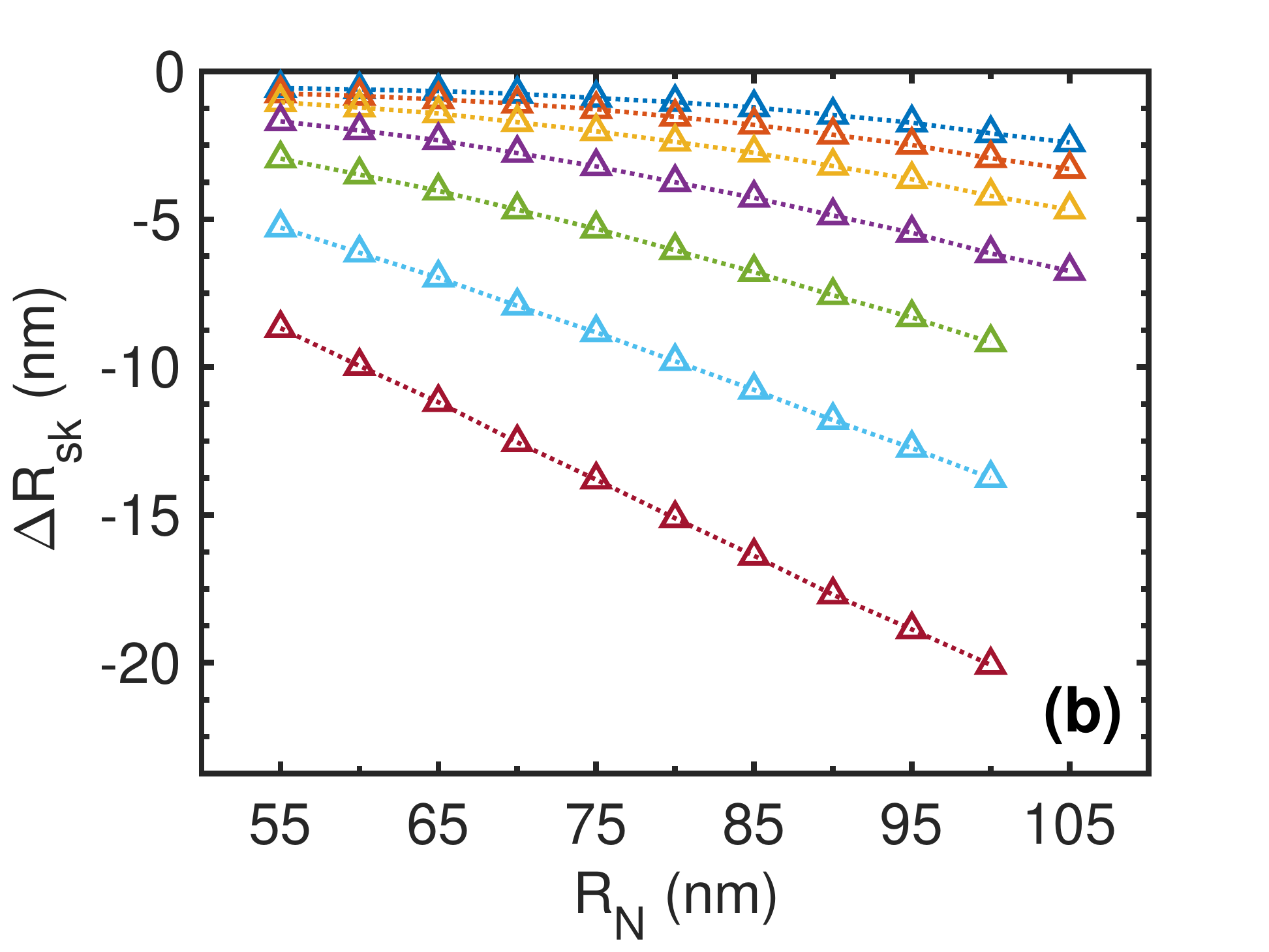}
\includegraphics[width=.32\textwidth]{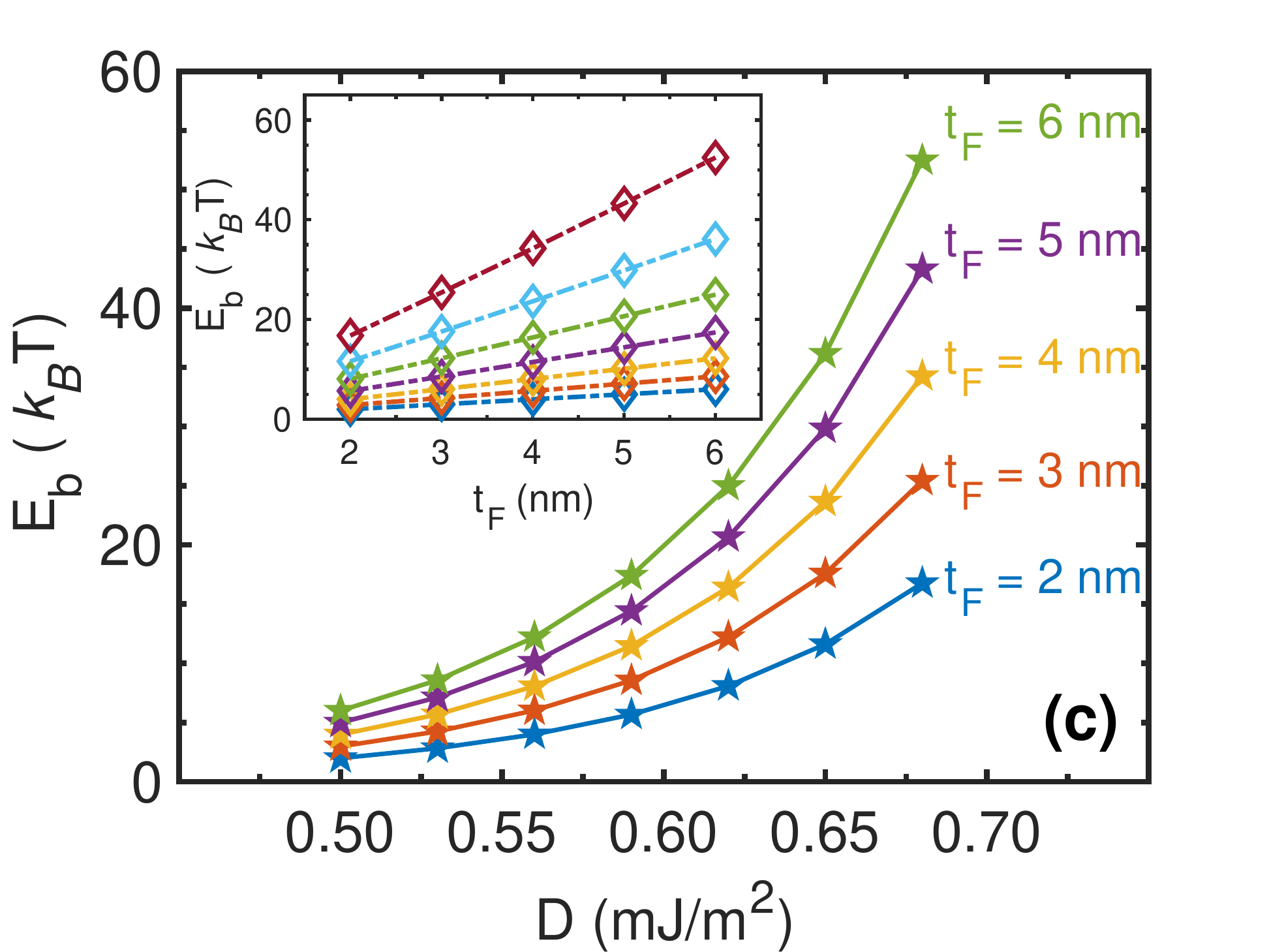}
\caption{Material parameters dependence of the energy barrier, $E_b$. Effect of notch size (radius $R_N$) on (a) $E_b$, and (b) $\Delta R_{sk}$, both for varying DMIs. We see that a larger $R_N$ leads to a larger $|\Delta R_{sk}|$ that corresponds to a higher barrier. (c) Thickness dependence of $E_b$ in a racetrack ($R_N = 100~nm$). The inset shows that $E_b$ increases linearly as a function of racetrack thickness, $t_F$ for any specific $D$. The linear variation of $E_b$ vs $t_F$ is consistent with the overall uniform cylindrical shape of the skyrmion at ultrathin limit. The color code to represent DMI variations in (a), (b), and inset of (c) are the same.}  
\label{fig2}
\end{figure*}
%%%%%%%%%%%%%%%%%%%%%%%%%%%%%%%%%%%%%%%%%%%%%%%%%%%%%%%%%%
%%%%%%%%%%%     Fig 3       %%%%%%
%%%%%%%%%%%%%%%%%%%%%%%%%%%%%%%%%%%%%%%%%%%%%%%%%%%%%%%%%%
\begin{figure*}[!htbp]
\includegraphics[width=.32\textwidth]{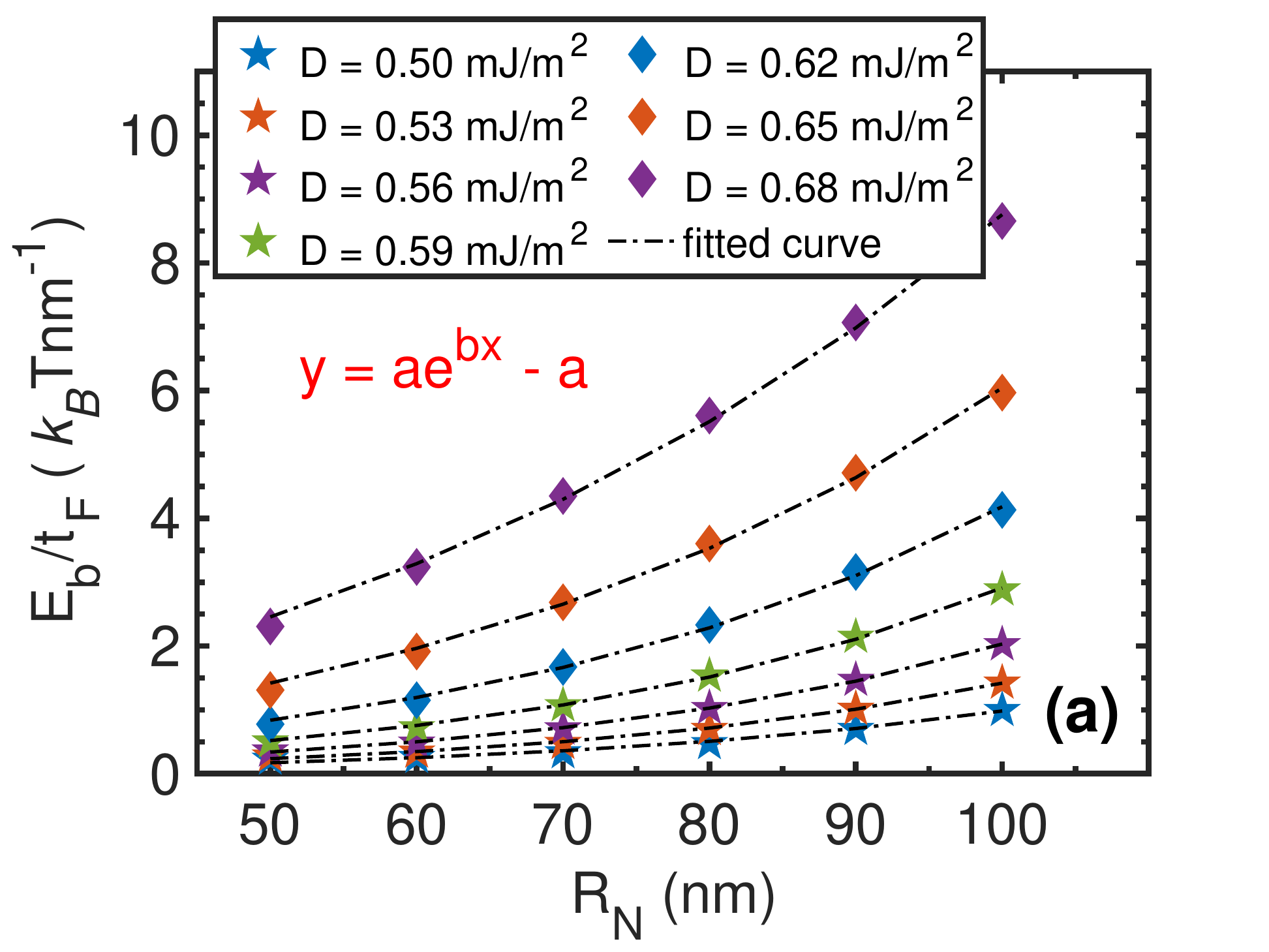}
\includegraphics[width=.32\textwidth]{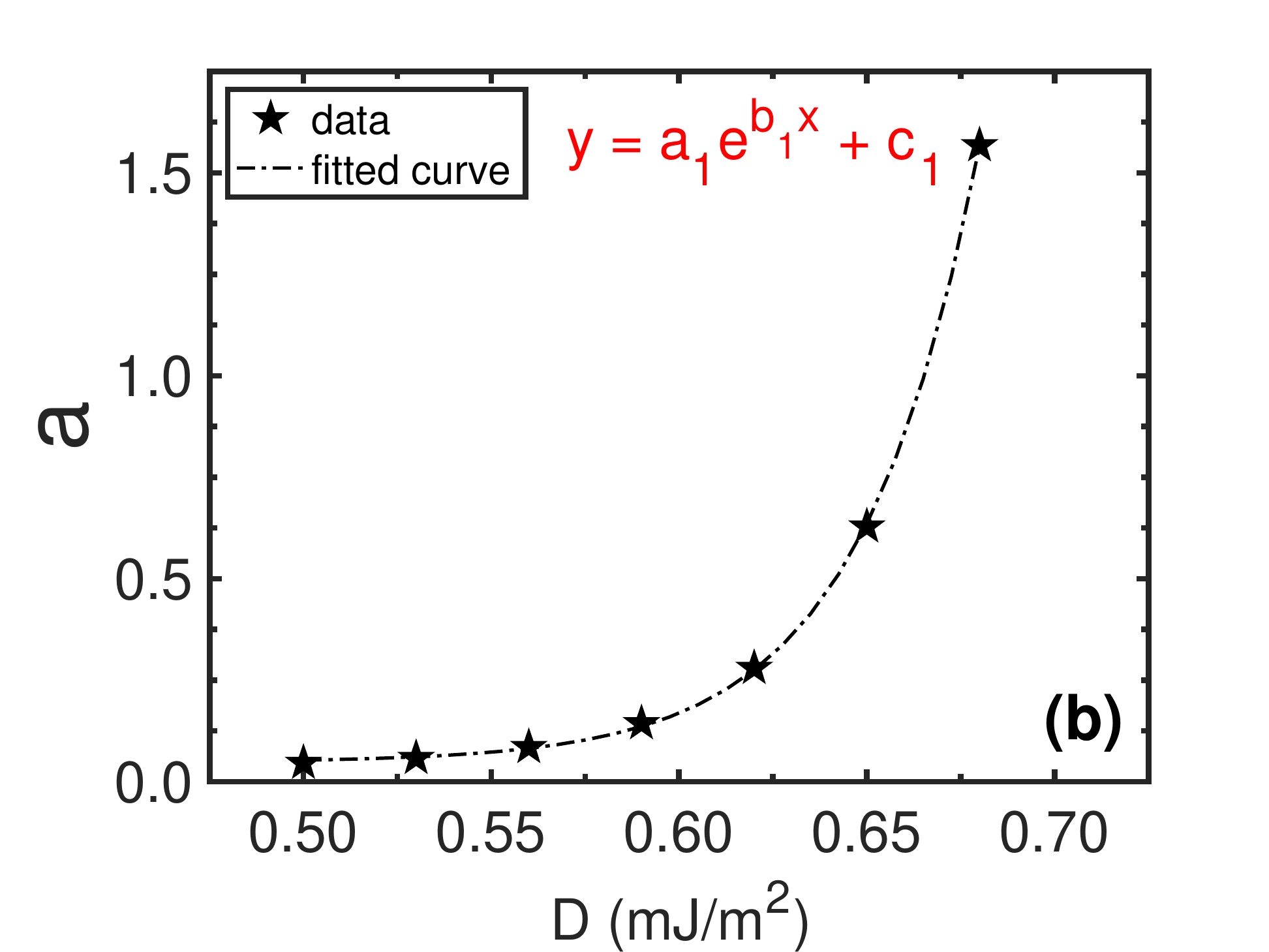}
\includegraphics[width=.32\textwidth]{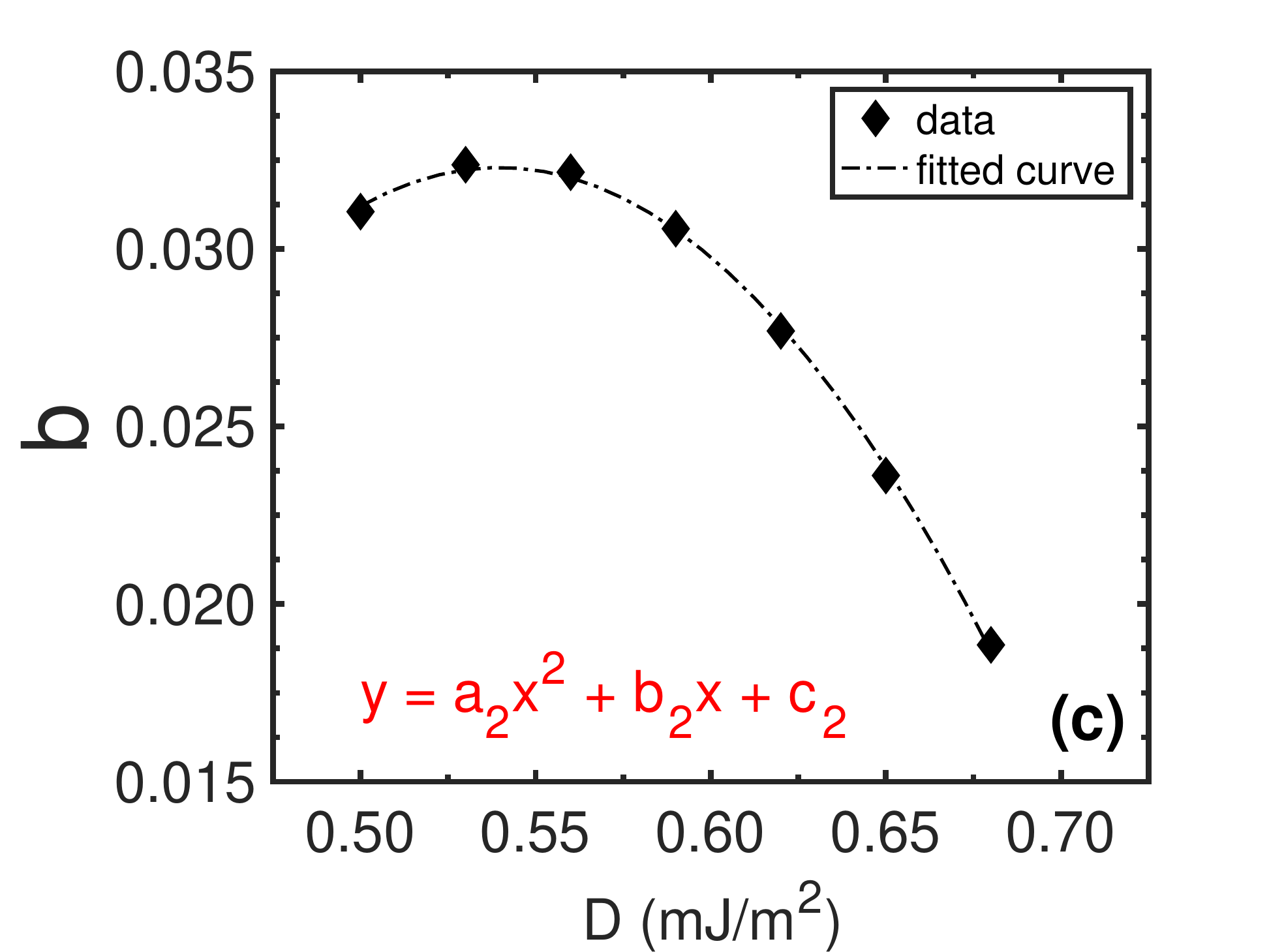}
    \caption{(a) Fitting of $E_b$ (normalized by $t_F$) with respect to $R_N$. (b), (c) The relation between the fitting constants obtained from (a) and $D$. The red colored texts in each graph represent the fitting function.}
    \label{fig3}
\end{figure*}
%%%%%%%%%%%%%%%%%%%%%%%%%%%%%%%%%%%%%%%%%%%%%%%%%%%%%%%%%%
%%%%%%%%%%%     Fig 4      %%%%%%
%%%%%%%%%%%%%%%%%%%%%%%%%%%%%%%%%%%%%%%%%%%%%%%%%%%%%%%%%%
\begin{figure*}[!htbp]
\includegraphics[width=0.85\textwidth]{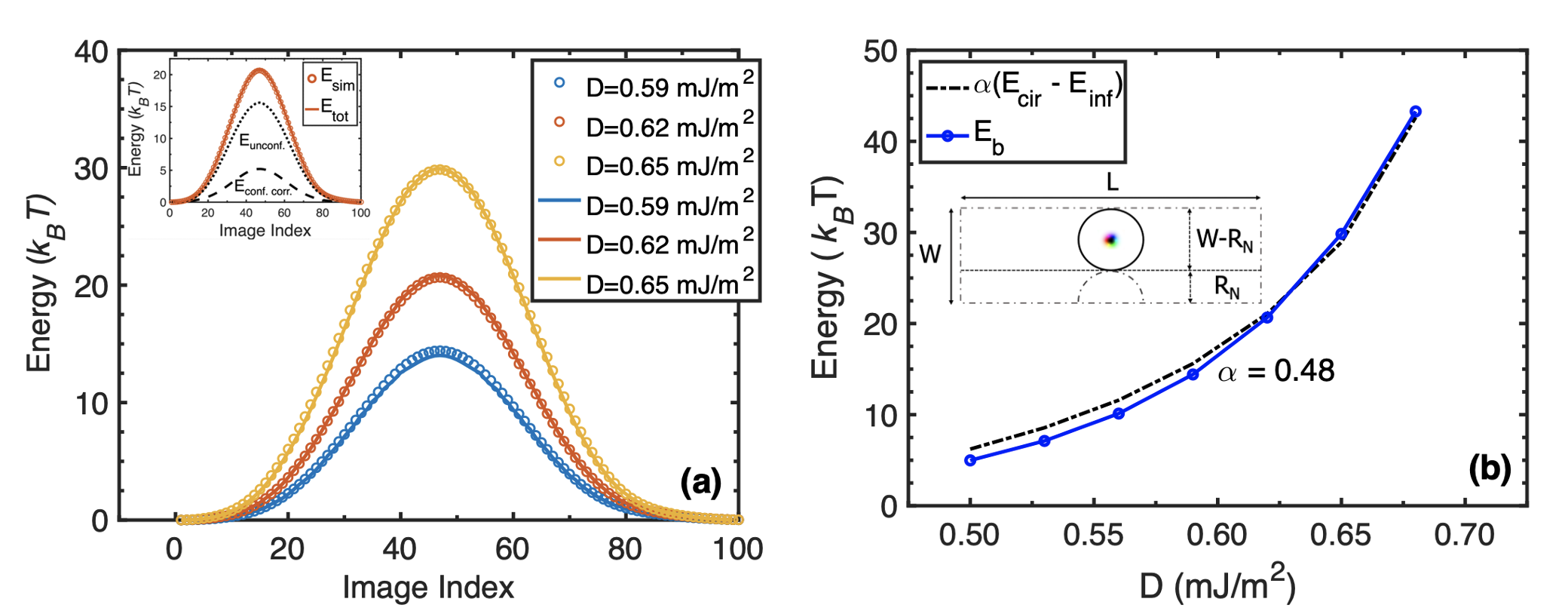}
    \caption{(a) Energy landscape of skyrmions in a racetrack ($R_N=100~nm$) from the MEP simulations (scatter circles) and analytical equations (solid curves)~\cite{skm_theory,Review2}. The analytical equations include the energetics of $2\pi$ skyrmions on an infinite plane plus a phenomenological confinement correction. (b) Explanation of the mismatch between simulated and analytical (calculated using equation~\ref{eneqns}) data in a racetrack ($R_N=100~nm$). A skyrmion confined above the pinning site has the energy of a skyrmion confined in a circle of diameter $W-R_N$ above the notch minus that of an unconfined infinite plane skyrmion. Including only one fitting term can describe $E_b$ with good accuracy for varying $D$ values. The black circular region above the notch in the inset schematic represents the simulation geometry for the confined case.}
    \label{fig4}
\end{figure*}
\begin{figure}[!htbp]
    \centering
    \includegraphics[width=0.90\linewidth,scale=0.5]{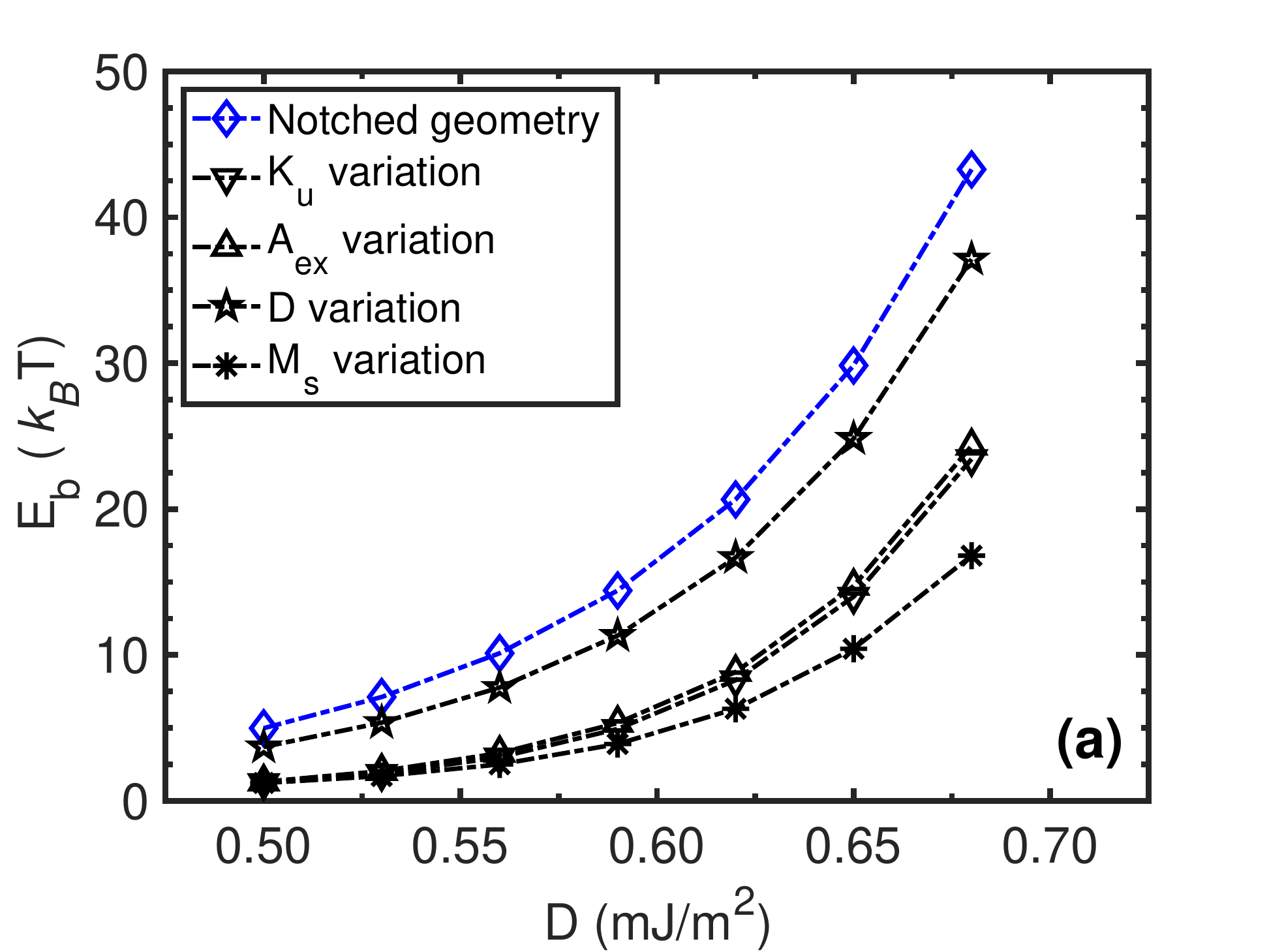}
     \includegraphics[width=0.90\linewidth,,scale=0.5]{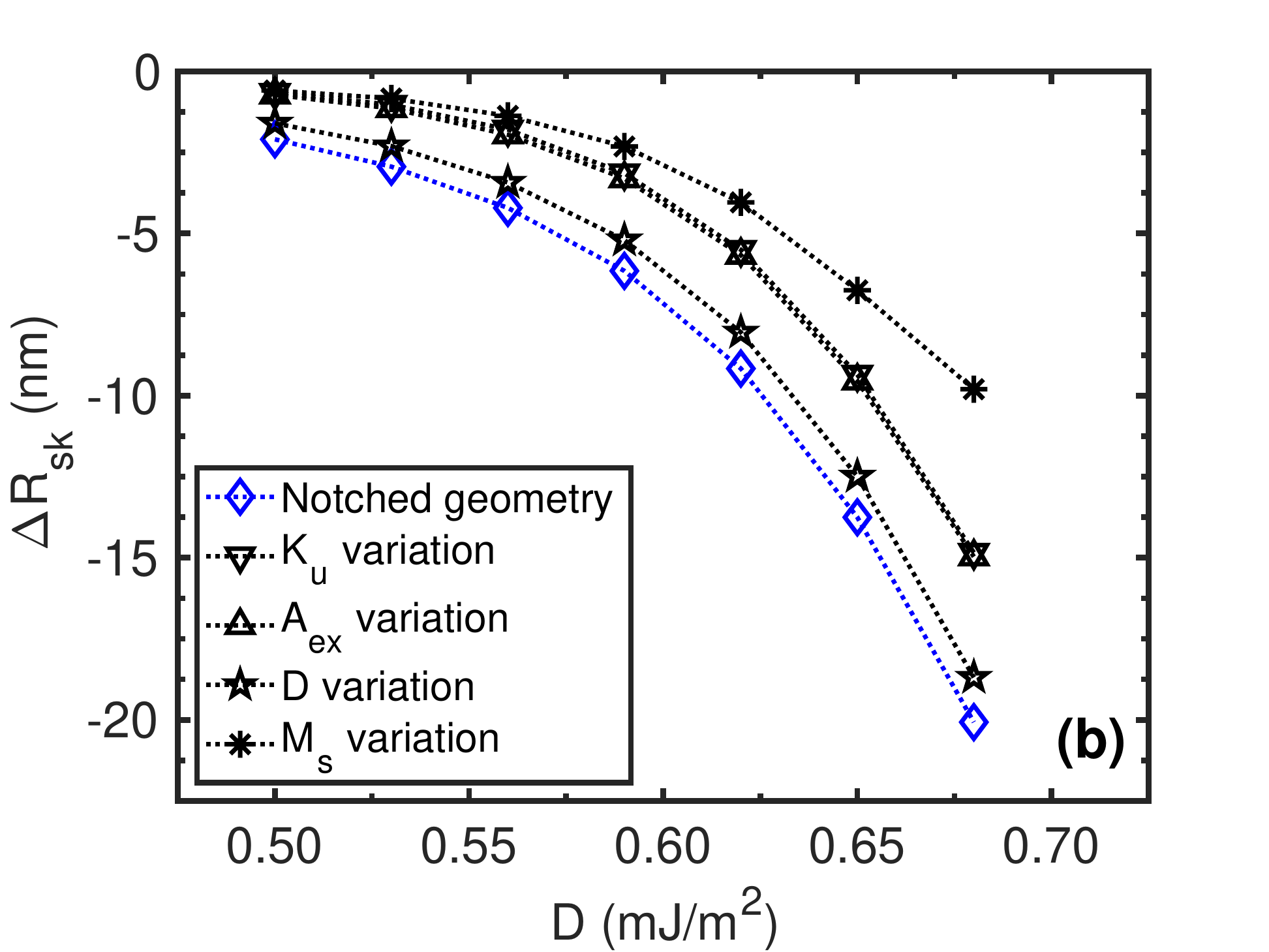}
    \caption{(a) Energy barrier for different pinning sites in a $5~nm$ thick racetrack. For all the cases, the pinning site is a semi-circular region of $100~nm$ radius. $K_u$ and $A_{ex}$ of the pinning site are $2$ times higher, and $D$ and $M_s$ are $10$ times lower than the rest of the track region. We choose the ratio that gives the highest $E_b$. (b) Change in skyrmion radius corresponding to the $E_b$ in (a), shows a proportional relation between $E_b$ and $|\Delta R_{sk}|$.}
    \label{fig5}
\end{figure}

Figure~\ref{fig1}(a) shows the schematic of a racetrack with notched geometry. We create the semi-circular notch (radius $R_N$) by removing materials from the racetrack. The snapshots represent different positions of the skyrmion trajectory (referred to as `image index') along the racetrack as it moves from one side to the other side of the notch. Figure~\ref{fig1}(b) shows the total energy obtained from the MEP calculations for a racetrack with notch radius $100~nm$ and thickness $5~nm$ (recall the racetrack width is $200~nm$). The zero for energy is set as the energy of the first image index in the simulation domain shown in Fig.~\ref{fig1}(a) top. We vary the DMI, $D$ from $0.50~mJ/m^2$ to $0.68~mJ/m^2$ and find energy barriers range from $\sim 5$ to $45$ $k_BT$. We find that the energy barrier results from the change in the skyrmion radius, $\Delta R_{sk}$ as the skyrmion passes past the notch. We conjecture that the reduction of the skyrmion size in squeezing through the constriction produces the energy barrier.\\
\indent Figure~\ref{fig1}(c) shows a series of $\Delta R_{sk}$ corresponding to the energy plots shown in Fig.~\ref{fig1}(b). As the $D$ increases, for a specific exchange and anisotropy, the skyrmion size gets bigger, making it harder to squeeze through and generating a higher energy barrier. The positional lifetime of the skyrmion is often described using an Arrhenius form $\tau = f_0^{-1} e^{E_b/k_BT}$ where $f_0$ is the attempt frequency and $E_b$ is the height of the gaussian energy profile (energy barrier). Approximately an $E_b$ of $30~k_BT$ ($35~k_BT$) will provide positional lifetime in seconds (days) for $f_0=10^{10}~Hz$. $45~k_BT$ energy barrier will give a lifetime in years.\\
\indent We calculate the MEP for various notch sizes and demonstrate the impact on the energy barrier in Fig.~\ref{fig2}(a) for $R_N = 55 - 105~nm$. We find that for all the DMIs, the energy barrier increases as $R_N$ increases, by reducing the size of the skyrmion, consistent with Fig.~\ref{fig2}(b). As $R_N$ increases, the skyrmion size shrinks more, $|\Delta R_{sk}|$ gets larger, and the energy barrier increases proportionally. However, if we continue to increase $R_N$, at some point, the skyrmion starts annihilating as the region over the notch is insufficient to pass through and the skyrmion touches the notch boundary and the edge of the racetrack. For instance, from Fig.~\ref{fig2}(a), we can see that throughout the range of the DMIs ($D = 0.50 - 0.68~mJ/m^2$), skyrmions pass through without annihilation up to $R_N=100~nm$. For a larger notch, for example, $R_N=105~nm$, the skyrmion gets annihilated when $D$ is greater than $0.59~mJ/m^2$.\\
\indent We also vary the thickness $t_F$ of the racetrack for several DMIs, and find an increase in energy barrier height for a thicker racetrack for a specific $D$. Figure~\ref{fig2}(c) shows the thickness dependence of the energy barrier in a racetrack with $R_N = 100~nm$. We get an energy barrier of $\sim 45~k_BT$ for a $\sim 45~nm$ skyrmion ($D=0.68~mJ/m^2$) in a moderately thick ($5~nm$) racetrack, which ensures years long lifetime that makes the device suitable for storage class memory applications. The inset shows the linearity of $E_b$ as a function of $t_F$, which dictates that we can increase the energy barrier even further by increasing the thickness of the racetrack. Clearly, we can get a large enough energy barrier for smaller skyrmions as well in a thicker racetrack.\\
\indent To quantify $E_b$, combining the data we get by varying the $D$, $R_N$, and $t_F$, we come up with a fitted empirical equation, normalized by $t_F$. Fig.~\ref{fig3}(a) shows the simulated $E_b$ vs $R_N$ data are perfectly described by the fitted curves of the form $y=a(e^{bx}-1)$. We find that $a$ is an exponential function of $D$, while $b$ is a quadratic function of $D$, as shown in Fig.~\ref{fig3}(b) and \ref{fig3}(c) respectively. 
The final form of the equation is
    \begin{equation}
        E_b/t_F=a(e^{bR_N}-1)
        \label{equRn}
    \end{equation}
where $a$ and $b$ are related to $D$ as follows
\begin{eqnarray}
a &=& a_1e^{b_1D}+c_1\nonumber\\
b &=& a_2D^2+b_2D+c_2
\label{equD}
\end{eqnarray}
\noindent where the prefactor constants are material specific. For GdCo, $[a_1,~b_1,~c_1]=[6.874\times 10^{-10},~31.64,~0.04783$, and $[a_2,~b_2,~c_2]=[-0.6922,~0.7471~,-0.1693$]. The units of $E_b$, $t_F$, $D$, and $R_N$ are in $k_BT$, $nm$, $mJ/m^2$, and $nm$ respectively. It is worth mentioning that the form of the equation (\ref{equRn}) is physically meaningful as it gives $E_b=0$ when $R_N=0$, which is expected. As energy of skyrmion increases linearly with $t_F$ for the limit of uniform cylindrical shape of skyrmion (inset of Fig.~\ref{fig2}(c)) at the limit of ultrathin films ($\lesssim 10~nm$), the $t_F$ is simply a scaling factor.\\
\indent We compare our simulated data with the analytical equation derived for skyrmions~\cite{skm_theory,Review2}. The different energy terms that give total energy equation of the skyrmions on an infinite plane are derived as~\cite{Review2,skm_theory}
\begin{eqnarray}
E_\mathrm{ex} &=& (2\pi A_{ex}t_F)\left(\frac{2R_{sk}}{\Delta} + \frac{2\Delta}{R_{sk}}N_{sk}^2\right)f_\mathrm{ex}(\rho)\nonumber\\
E_\mathrm{DMI} &=& -\left(2\pi R_{sk}t_F\right)\pi Df_\mathrm{DMI}(\rho)\nonumber\\
E_\mathrm{ani} &=& \left(4\pi K_u t_F\right)R_{sk}\Delta f_\mathrm{ani}(\rho)
\label{eneqns}
\end{eqnarray}
where $R_{sk},~\Delta,~N_{sk}$ are skyrmion radius, domain wall width, and skyrmion winding number respectively. The form factors for small size, obtained by fitting numerical simulations, are given by~\cite{Review2,skm_theory}
\begin{eqnarray}
f_\mathrm{ex}(\rho) &\approx& \left[1+1.93\frac{\rho\left(\rho-0.65\right)}{\rho^2+1}e^{\displaystyle -1.48(\rho-0.65)
}\right]\nonumber\\
f_\mathrm{ani}(\rho) &\approx&  \left[1-\frac{1}{6\rho}e^{\displaystyle -\rho/\sqrt{2}
}  \right]
\nonumber\\
f_\mathrm{DMI}(\rho) &\approx& \left[N_{sk}+\frac{1}{2\pi \rho}e^{-\displaystyle\rho} \right]
\label{eneqnsfit}
\end{eqnarray}
where $\rho=R_{sk}/\Delta$. Figure~\ref{fig4}(a) shows that our simulated energy profiles (scattered circles) are perfectly matched with the analytical equation (solid curves) that includes the energetics of skyrmions within $2\pi$ model on an infinite plane $E_{unconf.}=E_{ex}+E_{DMI}+E_{ani}$, plus a phenomenological confinement correction $E_{conf.~corr.}$ (see the dotted and dashed black curves in the inset of Fig.~\ref{fig4}(a) for $E_{unconf.}$ and $E_{conf.~corr.}$ respectively; adding up these two curves give the $E_{tot}$). $E_{conf.~corr.}$ can be expressed as
\begin{equation}
 E_{conf.~corr.} = Ae^{-{(\frac{x-B}{C})}^2}
\end{equation}
where $A$ is the magnitude, $B=47$, $C=20$, and $x$ is a placeholder for `image index'. $A$ is a quadratic function of $R_N$ and $D$, and can be expressed as
\begin{eqnarray}
A &=& pR_N^2+qR_N+r\nonumber\\
p &=& p_2D^2+p_1D+p_0\nonumber\\
q &=& q_2D^2+q_1D+q_0\nonumber\\
r &=& r_2D^2+r_1D+r_0
\label{equdiff}
\end{eqnarray}
where $[p_2,~p_1,~p_0]=[-0.0661,~0.0648,~-0.0142]$, $[q_2,~q_1,~q_0]=[11.8989,~-11.8985,~2.7966]$, and $[r_2,~r_1,~r_0]=[-134.0478,~100.5548,~-10.6283]$. We note that we use $N_{sk} =1$ and the value of $R_{sk}$ and $\Delta$ obtained from simulations while calculating the energy from the equation~(\ref{eneqns}).
However, equation~(\ref{eneqns}) alone fails to capture the simulated energy profiles because it assumes an unconfined planar geometry, while in our simulations, we use a confined geometry. We conjecture that our simulated energy barrier will be somewhere in between the skyrmion energy on an unconfined infinite plane and that for a confined circular region around the notch. To verify, we calculate the static energy of skyrmions both for a circle ($E_{cir}$) above the notch region with diameter $W-R_N$ (see the inset of Fig.~\ref{fig4}(b)), and for an unconfined infinite plane ($E_{inf}$). In Fig.~\ref{fig4}(b), we show the energy difference between $E_{cir}$ and $E_{inf}$ for a racetrack with $R_N = 100~nm$. We find that our simulated $E_b$ matches with $\alpha (E_{cir} - E_{inf})$ for the entire range of D, where $\alpha$ is a prefactor dependent on the racetrack geometry. We compare $\alpha (E_{cir} - E_{inf})$ and $E_b$ for other notch radii and find overall agreement while $\alpha$ varies as a function of notch radius.\\
\indent We explore alternate pinning mechanisms to compare the energy barrier among them. One common approach to introduce pinning sites is a local variation of the material parameters in a specific region in the racetrack~\cite{pinning,Sampaio}. In practical systems, the variation of material parameters can be achieved by naturally occurring and intentional defects, grain boundaries, composition and thickness gradient in the thin films, voltage gating, modulating the heavy metal layer, etc. We create the pinning sites by locally varying $K_u$, $A_{ex}$, $D$, and $M_s$. We vary one parameter at a time while the other parameters remain constant throughout the racetrack. Figure~\ref{fig5}(a) shows the energy barrier for different pinning sites, including the fully notched geometry for a $5~nm$ thick racetrack having a semi-circular pining site of $100~nm$ radius. It appears that a racetrack with a fully notched geometry produces the highest energy barrier compared to the rest, which attributes to the largest change in skyrmion radius while passing over the notch as shown in Fig.~\ref{fig5}(b). Additionally, notches are easier to create experimentally than controlling local variations of material parameters.\\
\indent We also calculate $E_b$ for another promising material Mn$_4$N for skyrmion-based spintronics applications~\cite{param_Mn4N,Gushi,Mn4N,Marco_Mn4N}. Mn$_4$N is a ferrimagnet and attractive for hosting small and speedy skyrmions~\cite{Marco_Mn4N}. In a $5~nm$ thick Mn$_4$N racetrack ($R_N = 75~nm$), we find an $E_b$ of $\sim 45~k_BT$ for a $\sim 40~nm$ skyrmion, while for GdCo with identical $R_N$, $t_F$, and $R_{sk}$, the $E_b$ is $\sim 22~k_BT$. The used parameters for Mn$_4$N are $A_{ex} = 15~pJ/m$, $K_u = 110~kJ/m^3$, $M_s = 105~kA/m$~\cite{param_Mn4N}. Our finding suggests that Mn$_4$N offers a higher $E_b$ than GdCo, which is mainly because of the higher exchange stiffness of Mn$_4$N. Needless to say that the $E_b$ can be further increased by tuning the $R_N$ and $t_F$ of the racetrack.
%%%%%%%%%%%%%%%%%%%%%%%%%%%%%%%%%%%%%%%%%%%%%%%%%%%%%%%%%%  
%%%%%%%%%%%     Fig 5       %%%%%%
%%%%%%%%%%%%%%%%%%%%%%%%%%%%%%%%%%%%%%%%%%%%%%%%%%%%%%%%%%
%%%%%%%%%%%%%%%%%%%%%%%%%%%%%%%%%%%%%%%%%%%%%%%%%%%%%%%%%%  
%%%%%%%%%%%     Fig 6       %%%%%%
%%%%%%%%%%%%%%%%%%%%%%%%%%%%%%%%%%%%%%%%%%%%%%%%%%%%%%%%%%
\begin{figure}[!htbp]
    \centering
    \includegraphics[width=0.90\linewidth]{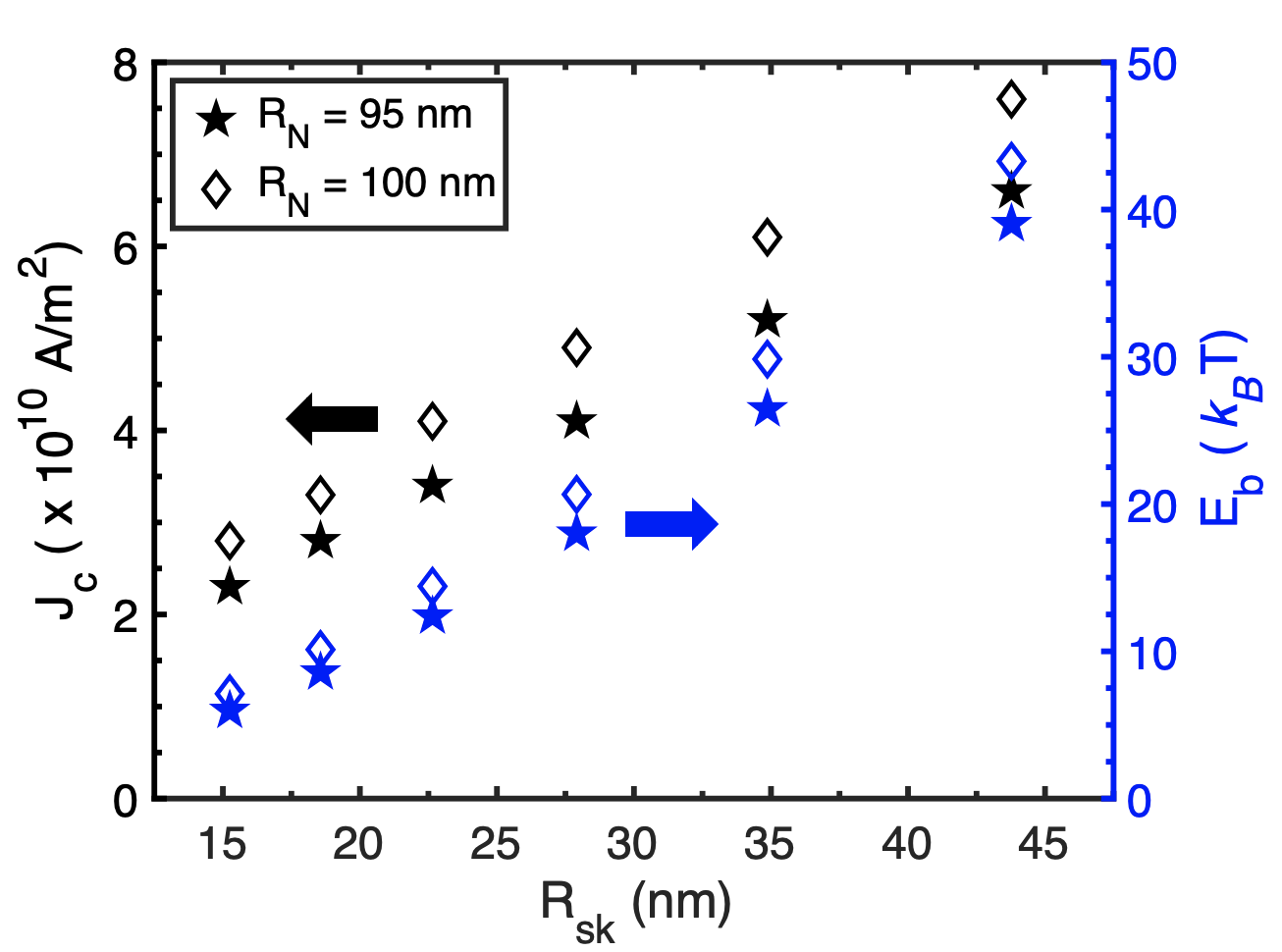}
    \caption{The critical current $J_c$ (black) to unpin the skyrmion and the corresponding $E_b$ (blue) for a $5~nm$ thick racetrack as a function of $R_{sk}$. We see a low unpinning current with a fairly large energy barrier. The arrows point each colored data to the corresponding $y$-axis.} 
    \label{fig6}
\end{figure}\\
\indent While a barrier is needed to hold the skyrmion in place, it is equally important to ensure that the critical current to depin the skyrmions is not too large, as that would cause unacceptable energy dissipation when integrated with the peripheral circuitry, not to mention random skyrmion annihilation, and even occasional unintended nucleation \cite{nucleation}. The energy barrier can be tuned by varying various knobs such as materials parameters and notch geometries. However, we need to optimize it to get a high enough hold time for the skyrmions yet require a moderate unpinning current.\\
\indent Figure~\ref{fig6} shows the unpinning current of racetracks with $95~nm$ and $100~nm$ notch radii, and $5~nm$ thickness. The current density distribution for the racetrack with the notch is calculated using COMSOL Multiphysics® \cite{COMSOL}. We use current pulses ranging from $8~ns$ to $25~ns$ to unpin the skyrmions. We find that bigger skyrmions need a shorter pulse and the critical current increases as $E_b$ increases. The energy barrier increases faster with radius than the critical current, which would help us to get a large enough barrier and a small enough unpinning current. We find moderate critical currents for large energy barriers. For instance, a $\sim 45~nm$ skyrmion can be unpinned with currents of $6.6 \times 10^{10}~A/m^2$ and $7.6 \times 10^{10}~A/m^2$ while the corresponding energy barriers are $\sim 40~k_BT$ and $\sim 45~k_BT$ respectively, which are orders of magnitude smaller than the critical current required to unpin domain walls~\cite{racetrack,Gushi}. Moreover, the obtained critical currents are significantly lower than the nucleation current ($> 10^{12}~A/m^2$) of the skyrmions in a constricted geometry~\cite{Iwasaki2013Oct,nucleation}, which prevents any unintended nucleation of skyrmions during the unpinning process.
\section{Conclusion} 
In summary, we demonstrated that skyrmion positional stability is achievable by creating notches along a racetrack. We presented quantitative analyses, backed by analytical equations for various material parameters, notch geometries, and racetrack thicknesses. An optimal combination of skyrmion size, notch radius, and thickness of the racetrack provides a large enough energy barrier ($\sim$ 45~$k_BT$) to achieve a positional lifetime of years for long-term memory applications. We found a moderately low minimum critical current to unpin the skyrmion ($\sim 10^{10}~A/m^2$), which is an essential aspect for low-power operations. These results provide critical design insights on skyrmionic racetracks, and potentially an argument for reliable, long-term skyrmion-based memory applications.
\section{Acknowledgments}
This work is funded by the DARPA Topological Excitations in Electronics (TEE) program (grant D18AP00009). The calculations are done using the computational resources from High-Performance Computing systems at the University of Virginia (Rivanna) and XSEDE. We thank Mark Stiles, Andrew Kent, Prasanna Balachandran, Mircea Stan, Geoffrey Beach, and Joe Poon for useful discussions.\\\\\\
\indent MG.M. and H.V. contributed equally to this work.
\bibliography{main,supporting}
\bibliographystyle{apsrev4-1}
\end{document}